# Chain-NN: An Energy-Efficient 1D Chain Architecture for Accelerating Deep Convolutional Neural Networks

Shihao Wang, Dajiang Zhou, Xushen Han, Takeshi Yoshimura
Graduate School of Information, Production and Systems, Waseda University, Kitakyushu, Japan
wshh1216@moegi.waseda.jp

*Abstract*— Deep convolutional neural networks (CNN) have shown their good performances in many computer vision tasks. However, the high computational complexity of CNN involves a huge amount of data movements between the computational processor core and memory hierarchy which occupies the major of the power consumption. This paper presents Chain-NN, a novel energy-efficient 1D chain architecture for accelerating deep CNNs. Chain-NN consists of the dedicated dual-channel process engines (PE). In Chain-NN, convolutions are done by the 1D systolic primitives composed of a group of adjacent PEs. These systolic primitives, together with the proposed column-wise scan input pattern, can fully reuse input operand to reduce the memory bandwidth requirement for energy saving. Moreover, the 1D chain architecture allows the systolic primitives to be easily reconfigured according to specific CNN parameters with fewer design complexity. The synthesis and layout of Chain-NN is under TSMC 28nm process. It costs 3751k logic gates and 352KB on-chip memory. The results show a 576-PE Chain-NN can be scaled up to 700MHz. This achieves a peak throughput of 806.4GOPS with 567.5mW and is able to accelerate the five convolutional layers in AlexNet at a frame rate of 326.2fps. 1421.0GOPS/W power efficiency is at least 2.5 to 4.1x times better than the state-of-the-art works.

## I. INTRODUCTION

Convolutional neural networks (CNN) have recently been a hot topic as they have led to great breakthrough in many computer vision tasks. Many CNN-based algorithms like AlexNet [1] and VGG [2] have proved to be stronger than conventional algorithms. Among these networks, convolutions usually account for the major part of the total computations in both training and testing phases. Further, there is a trend that CNN is towards deeper and larger for a better performance [4]. For example, ResNet [3] has recently proved a network with more than a thousand layers can further enhance the performance compared to a shallower network.

The increasing network scalability results in more computational complexities. It usually costs quite a long time to train deep CNNs and also challenges the design of real-time CNN applications. Therefore, CNN-specific accelerators are desired. The accelerators are expected to have not only a high reconfigurability to always achieve a high performance for various CNN parameters, but also the good energy and area efficiency for deployments on battery-limited devices.

To achieve these goals, researchers have been exploring efficient CNN accelerators on various platforms. The designs on CPU or GPU like [5] can achieve a high reconfigurability at the expense of high energy costs and usually suffer from the Von-Neumann bottleneck which limits their performance. Architectures on FPGA [6-7] or ASIC [8-15] are good candidates as these architectures can be totally customized to specific high-performance and energy-efficient needs. However, it usually involves high design complexities for supporting a high reconfigurability which will reversely affect the performance or efficiency. For example, [12] builds a 2D PE array, which has an on-chip network and inter-PE communications. Thus, an efficient architecture is desired to balance these metrics including the energy efficiency, for the increasing popularity of CNNs.

In this paper, we introduce our energy-efficient 1D chain architecture called Chain-NN for CNN accelerators, which contains the following contributions:

- We give a taxonomy of existing CNN accelerators to figure out their pros and cons.
- A novel energy-efficient 1D chain architecture is given. Its dedicated dual-channel architecture and column-wise scan input pattern provides energy-friendly reusability of input data to achieve 1.4TOPS/W power efficiency.
- 1D chain architecture of Chain-NN also supports a high reconfigurability by 84-100% of PE utilization rate for mainstream CNNs with low design complexities.
- Experimental results in TSMC 28nm show maximum 806.4GOPS throughput at 700MHz. Compared to state-of-the-art works, it achieves 2.5x to 4.1x power efficiency.

The rest of the paper is organized as follows. Sec. II briefly introduces CNN algorithms. In Sec. III, we give a taxonomy of existing CNN accelerators and introduce the features of our Chain-NN. Sec. III, we give a detailed discussion on the proposed 1D chain architecture. The experimental results are shown in Sec. IV. Finally, we conclude this paper in Sec. V.

## II. CNN BACKGROUND

Convolutional neural networks are trainable architectures where convolutional layers are the primary computational parts inside a network. Each layer receives input feature maps (ifmaps) and generates the output feature maps (ofmaps), which represent some particular features as shown in Fig. 1.

TABLE I
PARAMETERS OF A CNN CONVOLUTIONAL LAYER

| Parameter | Description |
|---|---|
| N | Batch Size |
| C/M | Number of ifmaps/ofmaps channels |
| H/E | Size of ifmaps/ofmaps |
| K | Size of kernels |

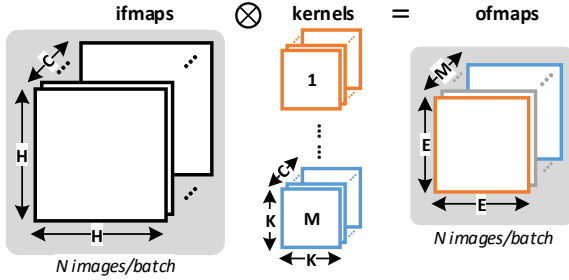

Fig. 1. A generalized example of computations in CNN convolutional layers.

We may regard ifmaps/ofmaps as 3D structures composed of multiple 2D channels. For each ofmaps channels, convolution is done between the 3D ifmaps and correspondent 3D kernels. Moreover, several group of 3D ifmaps compose of a batch, where the same group of kernels are shared. Therefore, convolutional layers can be regarded as 4D computations as shown in the Equation (1).

$$ofmaps[n][m][x][y] = bias[m] + \sum_{c=0}^{C-1}\sum_{i=0}^{K-1}\sum_{j=0}^{K-1} ifmpas[n][c][x+i][y+j] \times Kernel[m][c][i][j] \quad (1)$$

The meanings of indexes are shown in Table I. These parameters vary for different layers and different CNNs.

### III. HIGH LEVEL DESCRIPTION OF CHAIN-NN

*A. Taxonomy of Existing CNN Architectures*

We classify existing accelerator architectures into two categories according to how data are used inside processor cores. The way of data usage affects not only the reconfigurability, but also the required memory bandwidths (the amount of data movements). According to [16], the data movements of convolutional operands can be more expensive than ALU operations under the existing techniques. We give a detailed description and conclusion of the pros and cons of these two categories, followed by introducing how Chain-NN can improve over the existing works in Sec. III.B.

*1) Memory-centric Architectures*

As shown in Fig. 2(a), these architectures rely on the addressability of memories to achieve its reconfigurability for various CNN parameters. The processor core is simply a stacking of PEs. There is no data storage or data reuse paths inside the processor. A sophisticated central controller is necessary to decide when processor core should communicate with memories and which part of memories are accessed during the processing period.

These architectures can be reconfigured to adapt for various CNNs while the efficiency is sometimes unsatisfactory. They highly rely on memories for huge amount

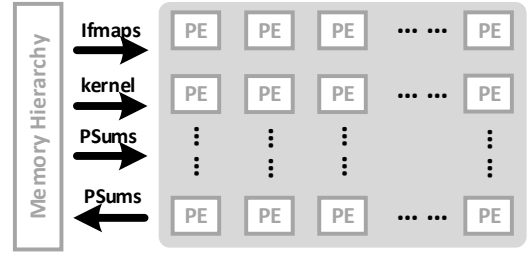

**(a) Memory-centric architecture** highly relies on the addressability of memories. Memory bandwidth requirement is high as all the data related to the processor core have to be interchanged with memories.

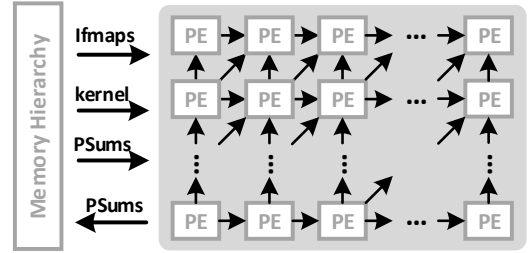

**(b) 2D spatial architecture** can reduce bandwidth requirement by reusing data among PEs. A complicated on-chip network is built to support inter-PE communication.

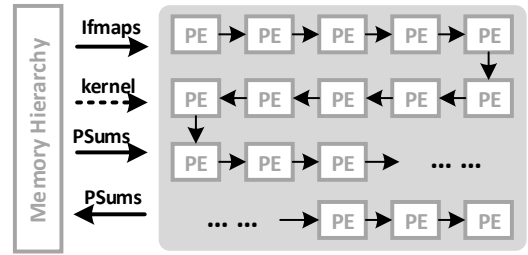

**(c) 1D chain architecture** organizes PEs as a chain architecture. The simplified inter-PE communication method can efficiently reduce not only the cost of controller, but also the amount of data movements. In this architecture, kernel is loaded only once per batch.

Fig. 2. An example of implementation using Adder-tree structure of 2D 3x3 convolution.

of data movements like [9,10,13]. Some achieve their reconfigurability by utilizing fewer PEs with fewer feeding data from memories, resulting in a waste of computing resources [8,15].

*2) 2D Spatial Architectures*

2D spatial architectures in Fig. 2(b) can reduce the data movements from memories by reusing them inside the processor core. Each PE can not only do basic computations, but also maintain a local controller to communicate with other PEs. PE usually contains local scratch pad memories to store data that are frequently used during a certain period. 2D spatial architectures divide the central controller in memory-centric ones and distribute them into each PE.

This solution can reduce the amount of data movements and have been employed in many works like [11,12]. However, their results show that the power and area cost of peripheral circuits for inter-PE communications can't be ignored. Moreover, this architecture has to consider the constraints in two dimensionalities for deployments, which sometimes limits its reconfigurability and scalability.

## B. High-level Description of Chain-NN

This paper introduces Chain-NN, a novel energy-efficient 1D chain architecture. As shown in Fig. 2(c), the PEs are organized as a chain architecture. Chain-NN is controlled by a finite state machine which changes its states according to a specific dataflow. The execution procedure is like this: 1) The finite-state machine is initialized to specific CNN parameters. 2) It starts to load related kernels into the processor core. 3) The ifmaps are continuously streamed into Chain-NN and convolution results are calculated. This paper mainly focuses on the design of processor core of Chain-NN and we leave the design exploration of memory hierarchy as the future work.

We conclude that Chain-NN has following advantages. First, it has a good energy efficiency compared to the others. It has a chain of dedicated dual-channel PEs which transfer data to adjacent PEs for data reuses with low control complexities. This not only reduces the total amount of memory accesses, but also shortens the distance of data fetching from on-chip SRAM level to PE level. Secondly, it has a high reconfigurability to be capable of achieving a high performance for various CNN parameters. This is due to its one dimensional organization of PEs, whose constraints are greatly relaxed compared to the 2D spatial architectures. An instantiation of our Chain-NN has proved to achieve an 84-100% PE utilization ratio considering the mainstreaming CNN parameters. Finally, this architecture involves fewer overheads when scaled up to a higher parallelism or clock frequency. This feature not only brings efficiencies of both power and area, but also makes the architecture flexible in matching various demands of designers.

## IV. CHAIN-NN: 1D CHAIN ARCHITECTURE

### A. 1D Chain Architecture

1D chain architecture is the processor core of Chain-NN. It consists of numbers of cascading PEs. A group of adjacent $K^2$ PEs forms a systolic primitive for convolution computations. An example of their relationship is shown in Fig. 3. We will give a detailed description on the 1D primitives in Sec. IV.B and the dual-channel PE architecture in Sec. IV.C.

In Fig. 3, the PE chain is cut into 1D primitives according to the kernel size. The upper part shows the case where each 1D primitive contains 9 PEs (K=3) while the lower refers K=2 case. The first and last PE in a primitive involve the ports to communicate with memory hierarchy, thus a set of primitive ports are attached.

The peak throughput of accelerator is proportional to the number of active PEs/primitives in the architecture. As a case study, we assume a systolic chain contains 576 PEs. The

TABLE II
THE NUMBER OF ACTIVE PE IN A 576-PE SYSTOLIC CHAIN

| Kernel Size | # of PEs of primitive | # of active primitives | # of active PEs | Efficiency |
|---|---|---|---|---|
| 3x3 | 9 | 64 | 576 | 100% |
| 5x5 | 25 | 23 | 575 | 99.8% |
| 7x7 | 49 | 11 | 539 | 93.6% |
| 9x9 | 81 | 7 | 567 | 100% |
| 11x11 | 121 | 4 | 484 | 84.0% |

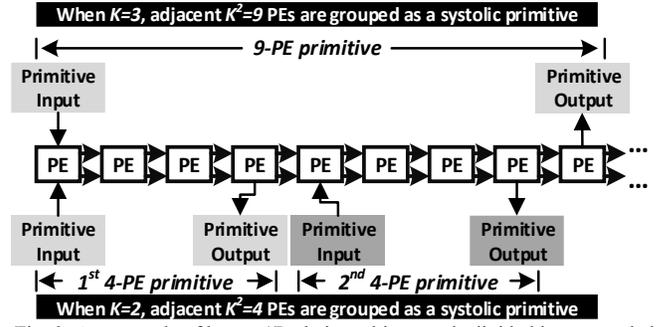

Fig. 3. An example of how a 1D chain architecture is divided into cascaded systolic primitives for various kernel size K.

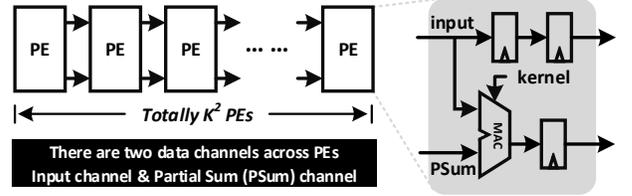

(a) Each pipeline stage forms a basic process engine (PE). The cascading PEs form a 1D systolic architecture for a 2D convolution operation.

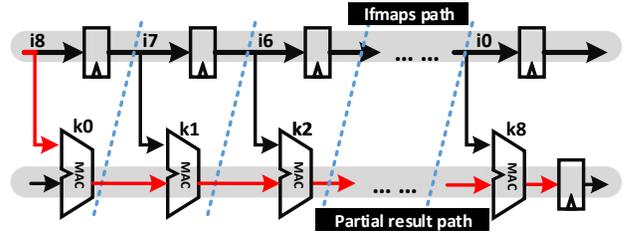

(b) Serial-in scheme eliminates the dependency between input bandwidth and kernel size. Pipeline is used to shorten the critical path delay.

Fig. 4. 1D systolic primitive for 2D convolution (3x3 case).

number of active PEs for mainstream kernel sizes is shown in Table II. The supported kernel sizes are not limited to these. Thanks to the 1D chain architecture, we are able to achieve at least 84% utilization rate of PEs.

### B. 1D Systolic Primitive for 2D Convolution

To support 1D chain architecture, each primitive for convolutions should also be designed as 1D implementation. This is done by pipelining a chain of multiply-accumulate operations (MAC) in Fig. 4 to form a systolic architecture like [16], which we call 1D systolic primitive. Notice that other pipelining schemes may produce more efficient architectures and we will discuss them in the future. In this paper, we mainly focus on how systolic primitives are employed inside the proposed 1D chain architecture for eliminating data locality.

Before that, we first present some design details about the 1D systolic primitives. It consists of a group of $K^2$ identical PEs which is shown in Fig. 4(a). The $K^2$ PEs are mapped to a convolutional kernel window. Thus, each PE is in charge of a 16-bit fixed-point MAC operation with a specific kernel weight.

Instead of reading the data in parallel, data are streamed in and go through every PE along the ifmaps path in Fig. 4(b). This guarantees an invariant input bandwidth requirement regardless of the kernel size K. Whenever a new pixel is streamed in at time t, previous (K-1) pixels during $[t-K^2+1, t-1]$ and this pixel constitute a 2D convolution window in ifmaps.

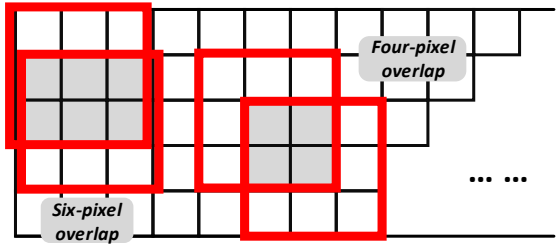

(a) Two convolutional windows have at most six overlapped pixels, so that three more pixels should be fetched in.

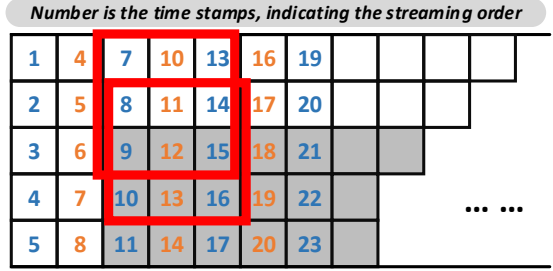

(b) Dual-channel PE has OddIF and EvenIF, in charge of odd and even columns respectively. Column-wise scan of convolutional windows guarantees $[t-K^2+1, t]$ pixels form a convolutional window since 9th cycle for any given $t$.

Fig. 5. The input order of ifmaps affects the PE throughput. Dual-channel PE architecture can achieve full utilization rate.

Note that kernels are stationary inside PEs so the convolutional window of ifmaps pixels during $[t-K^2+1, t]$ should be collocated exactly with the stationary kernel weights (matching issue). Sec. IV.C will introduce the proper input pattern and the proposed PE architecture to solve this matching issue.

Moreover, we can pipeline the critical path (red line) to shorten the output delay by making vertical cuts at each output of MAC (blue dotted line in Fig. 4(b)).

In conclusion, we call this architecture as a 1D systolic primitive, in which stationary kernel weights are stored inside each PE and input data serially flow through all the PEs. Each PE can communicate to adjacent PE. Thus, both the input data and the partial sums of MAC during convolution are passed and reused inside the systolic primitive, without accessing external memory. Meanwhile, it has a fixed input bandwidth requirement and a constant output delay regardless of the kernel size. Therefore, this 1D systolic primitive can not only reduce the memory accesses, but also provide a high scalability for various design demands.

*C. Dual-channel PE Architecture*

This part presents the dual-channel PE architecture and the column-wise scan input pattern. The dual channel means there are two data feed channels along the ifmaps path in Fig. 4(b). It can guarantee the high performance with the invariant input bandwidth of systolic primitives.

We first explain why a single channel based PE architecture can't fully utilize the computational resources of the 1D primitive. Let's still take K=3 as an example in Fig. 5(a). We have mentioned the matching issue that only if the order of $[t-K^2+1, t]$ ifmaps pixels accurately matches the stationary kernel window can the systolic primitive start to perform a convolution operation. However, inside the 2D ifmaps shown in Fig. 5(a), at most six pixels are overlapped for any two convolutional windows. It means that at least three cycles are required for fetching the rest three pixels due to the single channel (one ifmaps pixel per cycle). The 1D primitive has to be idle during this period. Therefore, this case shows that one-channel PE architecture can only achieve 1/K (33% in this case) of the peak throughput.

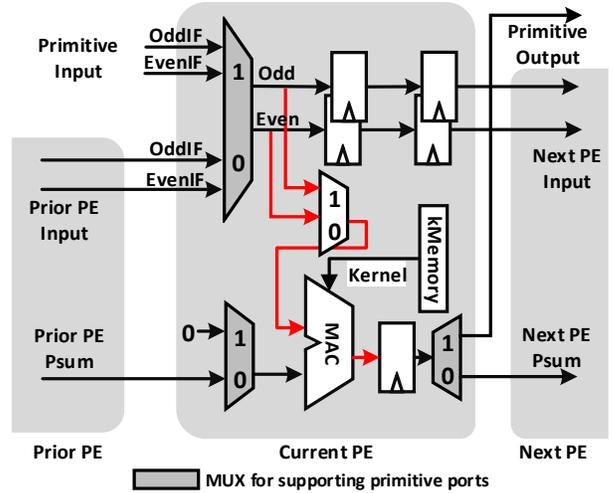

Fig. 6. Dual-channel PE architecture.

Therefore, we introduce the dual-channel PE architecture for this matching issue. Fig. 5(b) shows how we can benefit from it. We give a limitation that it requires K adjacent rows of ofmaps to be processed simultaneously. Then, (2K-1) rows of ifmaps (we refer this as input pattern in the following) are streamed in through two channels at the timestamps shown inside each pixel. Two channels are in charge of odd and even columns respectively. By doing this, we can find pixels of $[t-K^2+1, t]$ form a convolution window for any given t. Moreover, the pixel orders inside any window follow the column-wise scan order. We call this column-wise scan input pattern and it supports dual-channel PE can continuously start new convolutional operations after the initialization stage. Thus, we can 100% utilize the computational resources with the overheads of adding only one new ifmaps channel.

The dual-channel PE architecture is shown in Fig. 6. The dual channels (OddIF and EvenIF) inside each PE are in charge of transferring ifmaps into next PE. A multiplexer decided which channel's data is sent to MAC. Specifically, OddIF is in charge of odd columns (shown as blue numbers) while EvenIF only cares about the even columns (shown as orange numbers). EvenIF starts working after (K+1) cycle delay than OddIF. Meanwhile, a set of primitive input ports and function units (gray blocks in Fig. 6) are employed to support the systolic chain in Sec. IV.A. Inside each PE, there is a RegisterFile-based internal storage (kMemory) for storing the stationary kernels. The control of kMemory is designed to follow the dataflow. Moreover, we emphasize that we can easily pipeline the MAC path to shorten the critical path for a higher clock frequency, as shown by the red line in Fig. 6.

*D. Dataflow and Memory Hierarchy*

In a recent work [7], a detailed analysis on CNN memory-efficient dataflow and on-chip memory hierarchy have been discussed. We utilize their strategies for testing our proposed 1D chain architecture. The dataflow is modified to support the column-wise scan input pattern requirement as discussed in

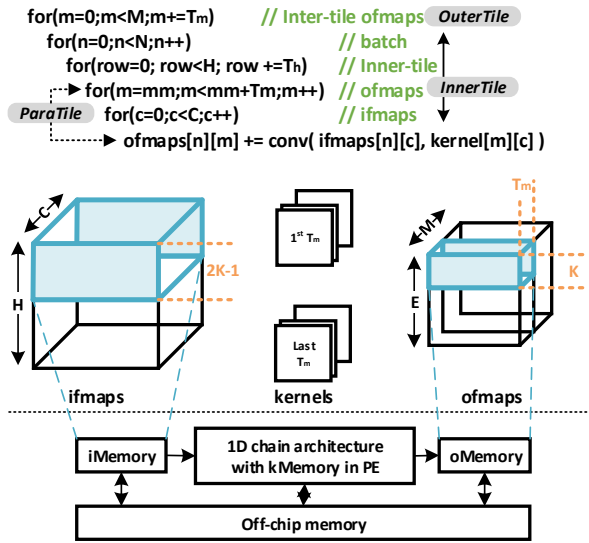

Fig. 7. The dataflow and memory hierarchy used for testing the performance of our 1D chain architecture.

Sec. IV.C. The detailed dataflow is shown by a loop structure in Fig. 7. The parallelism of Chain-NN unrolls the loop of ofmaps and the convolutions, marked as ParaTile. Meanwhile, two separate memories, iMemory and oMemory, form the memory hierarchy for data reusing among InnerTile.

## V. EXPERIMENTAL RESULTS

### A. Methodology

Chain-NN has been coded by SystemVerilog HDL and synthesized with TSMC 28nm HPC library under slow operation conditions (0.81V, 125°C) using the Synopsys Design Compiler. Layout is done by Encounter as shown in Fig. 8. We implemented a float-point-to-fix-point simulator which is integrated with MatConvnet for generating the test dataset including convolutional layers of pre-trained networks for MNIST, Cifar-10, AlexNet and VGG-16. Verification is done using ModelSim by both function simulation and post-synthesis simulation. The output results of hardware are checked with the simulator results on-the-fly. The power of the Chain-NN is analyzed by Power Compiler under typical operation conditions (0.9V, 25 °C). We generate switching activity interchange format (SAIF) by simulating the post-synthesis designs for power simulation.

### B. Performance

We instantiate a Chain-NN with 576 PEs, each of which is pipelined into three stages so that the critical path delay is reduced to 1.428ns (700MHz). Theoretically, this design can achieve a peak throughput of 806.4GOPS.

AlexNet is used to evaluate the realistic performance. We use totally 352KB on-chip memories, including kMemory in each PE, to support the data reuse in AlexNet based on the dataflow in Fig. 7. In detail, we implement 32KB for iMemory, 295KB for kMemory and 25KB for oMemory. In terms of kMemory, 295KB are averagely distributed into 576 PEs, resulting in a capable of 256 kernel weights per PE.

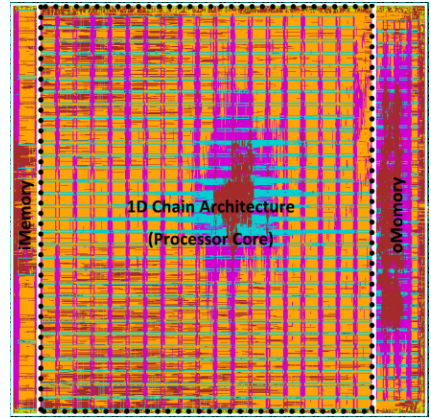

Fig. 8. Snapshot of the Layout.

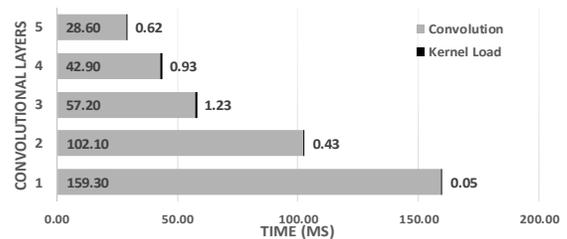

Fig. 9. The time distribution of five convolutional layers in AlexNet assuming the batch size is 128. Most of the time are cost by convolutional operations as kernel is loaded only once per batch.

TABLE IV
MEMORY COMMUNICATION BREAKDOWN WITH BATCH=4

| Layer / (MByte) | 1 | 2 | 3 | 4 | 5 | Total |
|---|---|---|---|---|---|---|
| DRAM | 9.0 | 5.5 | 4.3 | 3.4 | 2.3 | 24.5 |
| iMemory | 6.6 | 8.7 | 4.8 | 3.6 | 2.4 | 26.2 |
| kMemory | 15.4 | 17.8 | 37.2 | 27.9 | 18.6 | 116.8 |
| oMemory | 13.9 | 143.3 | 265.8 | 199.4 | 132.9 | 755.3 |

AlexNet contains five convolutional layers, including totally 666 millions of MACs per 227x227 input image. Fig. 9 presents the time distribution of five convolutional layers in AlexNet. For each batch, it costs 349.92ms while 3.25ms are spent for loading kernels. We emphasize that our architecture can benefit from a large batch size because we just load kernels once per batch, regardless of the batch size. Therefore, 326.2fps/275.6fps can be achieved for 128/4 batch sizes.

Table IV shows memory accesses and we emphasize Chain-NN greatly reduces the accesses for on-chip ifmaps and kernels accesses as well as off-chip accesses for all operands.

### C. Energy Efficiency

Energy efficiency is measured by the throughput per watt. Fig. 10 shows Chain-NN consumes 567.5mW and contributes 806.4GOPS, achieving a power efficiency of 1421.0GOPS/W.

In details, around 90% of the power consumption is from the 1D chain architecture including kMemory while only 10.55% is cost by the memory hierarchy. The power consumption of memories is greatly reduced in Chain-NN from two aspects. Firstly, we have shortened the length of most data movements from on-chip SRAM level to PE level. In detail, dedicated PEs with column-wise scan input pattern guarantee ifmaps are reused $K^2$ times averagely inside systolic primitives. This can reduce the data movements of ifmaps

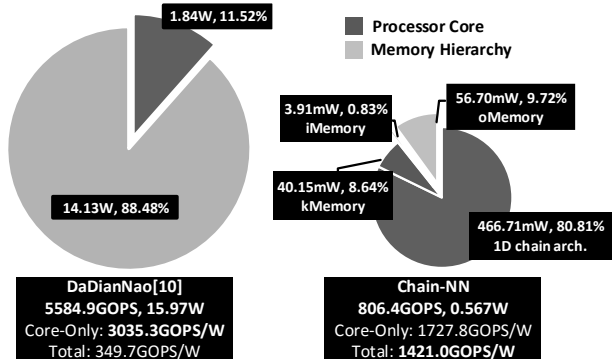

Fig. 10. The comparison of power efficiency with DaDianNao [10].

TABLE V
COMPARISON WITH STATE-OF-THE-ART WORKS

|  | DaDianNao[10] MICRO'14 | Eyeriss [12] ISCA'16 | Chain-NN |
|---|---|---|---|
| Technology | STM 28nm | TSMC 65nm | TSMC 28nm |
| Gate Count | N/A | 1852k | 3751k |
| On-chip Memory | 36MB eDRAM | 181.5KB SRAM | 352.0KB SRAM |
| Parallelism | 288x16 | 168 | 576 |
| Core Freq. | 606MHz | 250MHz | 700MHz |
| Power | 15.97W | 450mW | 567.5mW |
| Peak Throughput | 5584.9GOPS | 84.0GOPS | 806.4GOPS |
| Energy Eff. | 349.7GOPS/W | 245.6GOPS/W* | 1421.0GOPS/W |

*: If we scale this metric to 28nm to increase the core clock frequency, this work is expected to achieve **570.1GOPS/W** Energy Efficiency.

from SRAM to only ((2K-1)/K) times. Moreover, kernels are stored inside PEs so all the movements of kernels only happen inside PEs. Secondly, Chain-NN can reduce the maximum memory bandwidth requirement. For example, column-wise scan input pattern allows a kernel to be reused throughout the whole pattern (KE pixels). Therefore, kMemory can stay unchanged and has a very small activity factor of 1/KE. For example, the activity factor is only 2.22% for the third layers of AlexNet. Results also shows that kMemory consumes little power (40.15mW) through it has the largest 295KB size.

*D. Comparison with the State-of-the-art Works*

Two ASIC-based works [10][12] are chosen for comparison. DaDianNao [10] belongs to memory-centric architectures while Eyeriss [12] represents the 2D spatial ones.

Table V shows the comparison results. Our Chain-NN is at least 2.5x energy efficient. Fig. 10 gives the comparison with [10]. We roughly divided the designs into two parts, processor core and memory hierarchy. If only processor cores are considered, [10] can achieve around 3.0TOPS/W while Chain-NN is around 1.7TOPS/W. It is reasonable as we have reduced energy costs of data movements due to the analysis in Sec. V.C at the expense of increasing the complexity of the processor core. Therefore, Chain-NN can actually achieves 1421.0GOPS/W if we measure the whole chip (excluding the energy of off-chip DRAM accesses in Table IV).

Meanwhile, this work involves fewer area overheads when scaled up its parallelism. Considering the complexity of logic parts, Chain-NN only costs 6.51k/PE logic gates while [12] is 11.02k/PE. The reasons come from two aspects. First, 1D chain architecture has simplified data paths among PEs compared to the 2D spatial architectures. The reduced paths can still achieve a high throughput, which means that the paths in Chain-NN are utilized more efficiently. Secondly, the fewer memory bandwidth requirements can partially simplify the design complexity of memory hierarchy. These contribute to the 1.7 times area efficiency.

ACKNOWLEDGMENT

This work is supported by Waseda University Graduate Program for Embodiment Informatics (FY2013-FY2019).